\documentclass[proceedings]{JHEP3} 
\usepackage{epsfig,multicol}
\usepackage{amssymb}
\usepackage{graphics}
\def\beq{\begin{eqnarray}}    
\def\eeq{\end{eqnarray}}      

\newcommand{\OM}{\Omega_M}
\newcommand{\OL}{\Omega_{\Lambda}}

\newcommand{\OK}{\Omega_K}
\newcommand{\OZ}{\Omega_0}

\newcommand{\rc}{\rho_c}
\newcommand{\rM}{\rho_M}

\newcommand{\CC}{\Lambda}
\newcommand{\bCC}{\beta_{\Lambda}}

\PrHEP{}
\conference{International Workshop on Astroparticle and
High Energy Physics}

\title{A Friedmann-Lema\^{\i}tre-Robertson-Walker cosmological model with running $\Lambda$}

\author{
Ilya L. Shapiro $^{1}$, \speaker{Joan Sol\`a} $^{2}$\\
$^{1}\,$Universidade Federal Juiz de Fora, Minas Gerais, Brazil, E-mail: shapiro@fisica.ufjf.br\\
$^{2}\,$Dep. E.C.M.  Universitat de Barcelona, Diagonal 647,
E-08193, Barcelona, Spain, and C.E.R. for Astrophysics, Particle
Physics and Cosmology
\thanks{Associated with Instituto de Ciencias del
Espacio-CSIC.}, E-mail: sola@ifae.es}

\abstract{The idea of a variable dark energy has been entertained
many times in the literature and from many different points of
view. Quintessence is just a popular way to implement this idea
in recent times, but so far with little success. Another
possibility is to think of the cosmological term, $\Lambda$, as a
``running quantity'' much in the same way as the electromagnetic
coupling constant. However, the fact that $\Lambda$ is a
dimension-four parameter implies that it may obey a peculiar
renormalization group equation, which at low energies could be
dominated by ``soft decoupling'' contributions of the form
$\Lambda\sim H^2\,M_P^2\,$ stemming from physics near the Planck
scale. This value lies in the ballpark of the measurements from
CMB and high-z supernovae. A ``renormalized'' FLRW cosmology of
this kind may reveal itself as a sound, and testable, proposal for
a variable $\Lambda$ model within quantum field theory in curved
space time.}

\begin{document}
\providecommand{\href}[2]{#2}

\section{Introduction}

The Friedmann-Lema\^{\i}tre-Robertson-Walker (FLRW) cosmological
model\, \cite{Peebles} constitutes the standard paradigm of
present day cosmology. Its 4-curvature is determined from the
various contributions to its total energy-momentum tensor
density, namely in the form of matter energy density, radiation
pressure and cosmological constant:
\begin{equation}
R=8\,\pi\,G_{N}\,(\rho-3\,p+4\,\CC)\,. \label{totalR}
\end{equation}
The cosmological constant contribution to the curvature of
space-time is represented by the $\CC$ term. The latter enters the
original gravitational field equations in the form
\begin{equation}
R_{\mu \nu }-\frac{1}{2}g_{\mu \nu }R=-8\pi G_{N}\
\tilde{T}_{\mu\nu}\,, \label{EE}
\end{equation}
where $\tilde{T}_{\mu\nu}$ is given by $\tilde{T}_{\mu\nu}\equiv
T_{\mu\nu}+g_{\mu\nu}\,\CC (t)$, $T_{\mu\nu}$ being the ordinary
energy-momentum tensor associated to isotropic matter and
radiation. Here we entertain the possibility that $\CC$ is a
function of time\,\footnote{ While the homogeneous and isotropic
FLRW cosmologies do not allow spatial gradients of $\CC$, they do
not forbid the possibility that $\CC$ may be a function of the
cosmological time: $\CC =\CC (t)$.}. By the Bianchi identities, it
follows that $\CC$ is a constant if and only if the ordinary
energy-momentum tensor is individually conserved
($\bigtriangledown^{\mu}\,{T}_{\mu\nu}=0$). In particular, $\CC$
must be a constant if ${T}_{\mu\nu}$ is zero (e.g. during
inflation).

Modeling the expanding universe as a perfect fluid with velocity
$4$-vector field $U^{\mu}$, we have
\begin{equation}
T_{\mu\nu}=-p\,g_{\mu\nu}+(\rho+p)U^{\mu}U^{\nu}\,,
\label{Tmunuideal}
\end{equation}
where $p$ is the isotropic pressure and $\rho$ is the proper
energy density of matter.  Clearly the modified
$\tilde{T}_{\mu\nu}$ defined above takes the same form as
(\ref{Tmunuideal}) with $\rho\rightarrow
\tilde{\rho}=\rho+\CC\,,\ \ \ p\rightarrow \tilde{p}=p-\CC$. With
this generalized energy-momentum tensor, and in the FLRW metric
($k=0$ for flat, $k=\pm 1$ for spatially curved, universes)
\begin{equation}\label{FLRWm}
  ds^2=dt^2-a^2(t)\left(\frac{dr^2}{1-k\,r^2}
+r^2\,d\theta^2+r^2\,\sin^2\theta\,d\phi^2\right)\,,
\end{equation}
the gravitational field equations boil down to the
Friedmann-Lema\^{\i}tre equation
\begin{equation}
H^{2}\equiv \left( \frac{\dot{a}}{a}\right) ^{2}=\frac{8\pi\,G_N }{3}%
\left( \rho +\Lambda\right) -\frac{k}{a^{2}}\,,  \label{FL1}
\end{equation}
and the dynamical field equation for the scale factor:
\begin{equation}
\ddot{a}=-\frac{4\pi}{3}G_{N}\,(\tilde{\rho}+3\,\tilde{p})\,a=-\frac
{4\pi}{3}G_{N}\,(\rho+3\,p-2\,\CC)\,a\,. \label{newforce3}
\end{equation}
The physical CC is the parameter $\Lambda$ involved in the
equations (\ref{FL1}) and (\ref{newforce3}).  As these equations
are used to fit the cosmological data\,\cite{p99,Riess98}, the
parameter $\CC$ involved in them is to be considered the physical
(observed) value of the cosmological constant. A first integral of
the system (\ref{FL1}) and (\ref{newforce3}) is given by
\begin{equation}\label{Bronstein}
\dot{\CC}+\dot{\rho}+3\,H\,(\rho+p)=0\,,
\end{equation}
which shows that a time-variable CC cosmology allows transfer of
energy from matter-radiation into vacuum energy, and vice versa.
Since the radiation pressure is negligible at present,
Eq.\,(\ref{FL1}) can be rewritten as the following exact sum rule
for the cosmological parameters:
\begin{equation}\label{sumrule0}
\OM^0+\OL^0+\OK^0=1\,,
\end{equation}
 with
\begin{equation}
\OM^0\equiv \frac{\rM^0}{\rc^0}\,,\ \ \ \OL^0\equiv \frac{\CC_0
}{\rc^0}\,,\ \ \ \OK^0\equiv \frac{-k }{H_0^2\,a_0^2}\,,
\label{cpparameters}
\end{equation}
$\rM^0$ and $\CC_0$ being the matter density and cosmological
constant at our time, and
\begin{equation}
\rc^0 \equiv \frac{3\,H_{0}^{2}}{8\,\pi \,G_{N}} \simeq
10.6\,h_{0}^{2}\ GeV/m^{3}\simeq \left( 3.0\,\sqrt{h_{0}}\times
10^{-3}\,eV\right) ^{4} \label{rhocrit}
\end{equation}
is the present value of the critical density (equivalent to a few
protons per cubic meter). Here the dimensionless number
$h_{0}\sim 0.7\pm 0.1$\ sets the typical range for today's value
of Hubble's constant $H_{0}\equiv \left(
\dot{a}/{a}\right)_0=100\;(Km/sec\, Mpc)\;h_{0}$.

What is the nature of the total energy density of the universe?
On the one hand the baryonic contribution to the total matter
content $\rM$ (a scanty $\Omega_B\sim 5\%$ of the critical
density, derived from nucleosynthesis calculations) is far
smaller than the total amount of matter detected by dynamical
means \cite{Peebles,CMBR,WMAP}, namely $\OM^0\sim 30\%$ of the
critical density. Therefore  the bulk of the matter content of the
universe is dark matter, and must be in the form of an unknown
kind of cold (non-relativistic and non-baryonic) invisible
component. Significant amounts of hot (relativistic) dark matter
are excluded because it would not fit with the models of
structure formation (particularly at small scales). So the
radiation part at present boils down to an insignificant (few per
mil) fraction of neutrinos (yet comparable to the total amount of
luminous matter that we see!) plus an even more negligible
contribution of very soft photons (the relic CMB radiation)
entering at the level of one ten-thousandth, at most, of the
critical density today\,\cite{CMBR,WMAP}.

On the other hand, the astrophysical measurements tracing the rate
of expansion of the universe with high--z Type Ia supernovae
\cite{p99, Riess98} indicate that $\OL^0\sim 70\%$ of the
critical energy density of the universe is cosmological constant
(CC) or a dark energy candidate with  a similar dynamical impact
on the evolution of the expansion of the universe. Specifically,
the CC value found from Type Ia supernovae at high z is:
\begin{equation}\label{CCvalue}
\CC_0=\OL^0\,\rc^0\simeq 6\,h_0^2\times 10^{-47}\,GeV^{4}\,.
\end{equation}
Independent from these supernovae measurements, the CMB
anisotropies\,\cite{CMBR}, including the recent data from the WMAP
satellite,  lead to $\OZ=1.02\pm 0.02$\,\cite{WMAP}. As a first
observation, it is obvious that this result leaves little room for
our universe to be spatially curved ($\OK\leq 2\%$), and indeed it
suggests that our universe is spatially flat, as expected from
inflation\,\cite{Guth}. As a second observation, when combining
this result with the dynamically determined value (from clusters
of galaxies) of the matter density (viz. $\OM^0\simeq 30\%$), the
bookkeeping in Eq. (\ref{sumrule0}) leads us to an outstanding
conclusion: the rest of the present energy budget (a large gap of
order $70\%$ of the critical density) must be encoded in the
parameter $\OL^0$. Hence the CMB measurements and the high-z
supernovae data lead to a conclusion in concordance with the value
of $\CC$, if one accepts the data on $\OM^0$ from clusters. As
mentioned, the value (\ref{CCvalue}) corresponds to the parameter
entering the classical cosmological equations (\ref{FL1}) and
(\ref{newforce3}). Therefore the situation seems to be consistent
from the experimental point of view\,\footnote{See, however,
Ref.\,\cite{Subir} for a more critical point of view on the
present ``Concordance Model''.}

What about the theoretical situation? In Quantum Field Theory
(QFT) we have long expected that the vacuum fluctuations should
induce a non-vanishing value for $\CC$ \cite{zeldo}, and the
question is whether in realistic QFT's we have a prediction for
$\CC_0$ in the ballpark of the measured value (\ref{CCvalue}).
Sadly, the answer is no. For, in the context of the Standard
Model (SM) of electroweak interactions, this measured CC should
be the sum of the original vacuum CC in Einstein's equations,
 $\CC _{vac}$, and the induced
contribution from the vacuum energy of the quantum fields:
\begin{equation}
\CC=\CC _{vac}+\CC _{ind}\,. \label{lambdaphys}
\end{equation}
What is the expected value for $\CC _{ind}$ in the SM?  Let
$\Phi$ be the quantum Higgs field operator. In the ground (vacuum)
state of the SM, the vacuum expectation value (VEV) of $\,\Phi
^{+}\Phi \,$ will be denoted $\,<\Phi ^{+}\Phi
>\equiv \frac{1}{2}\phi ^{2}$, where $\,\phi \,$ is the mean
field associated to the field operator $\Phi$. The classical
potential associated to the mean field reads
\begin{equation}
V_{cl}=-\frac{1}{2}m_{\phi}^{2}\phi ^{2}+\frac{\lambda}{8}\phi
^{4}. \label{2a}
\end{equation}
Shifting the original field $\phi \rightarrow H^{0}+v$ such that
the physical scalar field $H^{0}$ has zero VEV one obtains the
physical mass of
the Higgs boson: $M_{H}=\sqrt{2}\;m_{\phi}$. Minimization of the potential (\ref{2a}%
) yields the spontaneous symmetry-breaking relation:
\begin{equation}
\phi =\sqrt{\frac{2m_{\phi}^{2}}{\lambda}}=v\,\,\,\,\,\,\mathrm{and} \,\,\,\,\,\,\,\lambda=%
\frac{M_{H}^{2}}{v^{2}}\,.  \label{5N}
\end{equation}
The VEV $<\Phi >\equiv v/\sqrt{2}$ gives masses to fermions and
weak gauge bosons through
\begin{equation}
m_{f_i}=h_{i}\,\frac{v}{\sqrt{2}}\,,\;\;\;M_{W}^{2}=\frac{1}{4}
\,g^{2}\,v^{2},\;\;\;M_{Z}^{2} =\frac{1}{4}\,(g^{2}\,+g^{\prime
2})\,v^{2}, \label{masses}
\end{equation}
where $h_{i}$ are the corresponding Yukawa couplings, and $g\,$ and $%
\,g^{\prime }\,$ are the $SU(2)_{L}$ and $U(1)_{Y}$ gauge
couplings. The VEV can be written entirely in terms of the Fermi
scale $M_{F}\equiv G_{F}^{-1/2}\simeq
293\,GeV$ as follows: $\,v=2^{-1/4}M_{F}%
\simeq 246\,GeV$. From (\ref{5N}) one obtains the following value
for the potential, at the tree-level, that goes over to the
induced CC:
\begin{equation}
\Lambda
_{ind}=<V_{cl}>=-\frac{m_{\phi}^{4}}{2\lambda}=-\frac18\,M_H^2\,v^2\,.
\label{nnn6}
\end{equation}
From the current LEP 200 numerical bound on the Higgs boson mass,
$\,M_{H}> 114.1\,GeV$ \cite{LEPHWG}, one finds $\,\,\,\left| \CC
_{ind}\right|> 1.0\times 10^{8}\,GeV^{4}$. Clearly, $\left| \CC
_{ind}\right|$ is $55$ orders of magnitude larger than the
observed CC value (\ref{CCvalue}). Moreover, the Higgs potential
gets renormalized at higher order in perturbation theory, and
therefore it is the value of the effective Higgs potential
$V_{\rm eff}$ what matters at the quantum level. These quantum
corrections by themselves are already much larger than
(\ref{CCvalue}). Finally, we also note that in general the induced
term may also get contributions from strong interactions, the
so-called quark and gluon vacuum condensates. These are also huge
as compared to (\ref{CCvalue}), but are much smaller than the
electroweak contribution (\ref{nnn6}).

Such discrepancy, the so-called
 ``old'' cosmological constant problem (CCP)
 \cite{weinRMP,CCRev}, manifests itself in the necessity of
enforcing an unnaturally exact fine tuning of the original
cosmological term $\CC _{vac}$ in the vacuum action that has to
cancel the induced counterpart $\,\CC _{ind}$ within a precision
(in the SM) of one part in $\,10^{55}$. This big conundrum has
triggered many theoretical proposals. On the first place there is
the longstanding idea of identifying the dark energy component
with a dynamical scalar field \cite{Dolgov,PSW}. More recently
this approach took the popular form of a ``quintessence'' field
slow--rolling down its potential\,\cite{Caldwell} and variations
thereof\,\cite{PeebRat}. The main advantage of the quintessence
models is that they could explain the possibility of an evolving
vacuum energy. This may become important in case such evolution
will be someday detected in the observations, although it is not
obvious that the models can be easily
discriminated\,\cite{Padmanabhan}. Furthermore, a plethora of
suggestions came along with string theory
developments\,\cite{wittenDM} and anthropic
scenarios\,\cite{antrop}. Other recent ideas have been put
forward, like the intriguing proposal of non-point-like gravitons
at sub-millimeter distances\,\cite{Sundrum2}, or the suggestion of
having multiple degenerate vacua\,\cite{Yokoyama}.

There are however other (less exotic) possibilities, which should
be taken into account. In a series of recent papers
\cite{JHEPCC1,cosm}, the idea has been put forward that already
in standard QFT it should not make much sense to think of the CC
as a constant, even if taken as a parameter at the classical
level, because the Renormalization Group (RG) effects may shift
away the prescribed value, in particular if the latter is assumed
to be zero.  Thus, in the RG approach one takes a point of view
very different from the quintessence proposal, as we deal all the
time with a ``true'' cosmological term. It is however a variable
one, and therefore a time-evolving, or redshift dependent:
$\CC=\CC (z)$. Although we do not have a QFT of gravity where the
running of the gravitational and cosmological constants could
ultimately be substantiated, a semiclassical description within
the well established formalism of QFT in curved space-time (see
e.g. \cite{birdav,book}) should be a good starting point. Then, by
looking at the CCP from the RG point of view\, \cite{Polyakov},
the CC becomes a scaling parameter whose value should be
sensitive to the entire energy history of the universe -- in a
manner not essentially different to, say, the electromagnetic
coupling constant.

The canonical form of renormalization group equation (RGE) for
the $\CC$ term at high energy is well known -- see e.g.
\cite{book}. However, at low energy decoupling effects of the
massive particles may change significantly the structure of this
RGE, with important phenomenological consequences. This idea has
been elaborated recently by several authors from various
interesting points of view \cite{JHEPCC1,cosm,Babic,Reuter03a}. It
is not easy to achieve a RG model where the CC runs smoothly
without fine tuning at the present epoch. In
Ref.\,\cite{Letter,IRGA03,CCfit} a successful attempt in this
direction has been made, which is based on the possible existence
of physics near the Planck scale. In the following we review the
main features and implications of this ``RG-cosmology''.

\section{Running cosmological constant}

It is well known that the effective action of vacuum takes the
form \cite{birdav,book}:
\begin{equation}
S_{vac}=\int d^{4}x\sqrt{-g}\,\Big\{
\,a_{1}R^{2}_{\mu\nu\alpha\beta}
+a_{2}R^{2}_{\mu\nu} + a_{3}R^{2} + a_{4}{\square} R-\frac{1}{16\pi G_{vac}}%
\,R - \Lambda_{vac}\Big\}\,,  \label{Svac}
\end{equation}
and at low energies the only relevant part for phenomenological
considerations is the Hilbert-Einstein piece
\begin{equation}
S_{HE}=-\int d^{4}x\sqrt{-g}\,\left( \,\frac{1}{16\pi G}%
\,R+\CC _{vac}\,\right) \,.  \label{HE}
\end{equation}
While the phenomenological impact of the higher derivative terms
in (\ref{Svac}) is negligible at low energy (namely, at current
times), the presence of the parameter $\Lambda_{vac}$ is as
indispensable as any one of these higher derivative terms to
achieve a renormalizable QFT in curved space-time\,\footnote{From
this sole observation it follows that quintessence models without
a $\Lambda$ term cannot be renormalizable theories in curved
space-time.}. Admittedly the vacuum CC itself, $\Lambda_{vac}$, is
not the physical (observable) value of the cosmological constant.
However, its presence is essential to make possible the matching
of the theoretical value of  $\CC$ in Eq.\,(\ref{lambdaphys})
with the physical measurement (\ref{CCvalue}). Here we do not
address the issue of why the two terms (induced and vacuum terms)
cancel with the desired precision (the ``old'' CCP). Rather, we
take it for granted, and then study the possibility that the
cosmological constant evolves according to a RGE whose flow
crosses the present instant of time with the value
(\ref{CCvalue}), and we then investigate whether such a RGE, when
extrapolated back to the past, gives testable and consistent
predictions. In particular, we verify whether the predicted value
of $\CC$ at the nucleosynthesis time does not disturb the basic
features of this momentous regime in the history of the universe.
Once this is guaranteed, we wish to check whether the predictions
of the RGE for nearby past times depart significantly from the
standard FLRW cosmology, and therefore whether we can find out
signs that can effectively discriminate among the two.  Finally,
the same RGE can give us a prediction for the future or destiny of
our universe.

Of course the first issue is to elucidate the structure of the
RGE for the $\CC$ term at any given energy regime. A first clue is
to recognize that this renormalization group equation is well-know
at high energy\,\cite{book,JHEPCC1}.  For particles $i=1,2,3,...$
of masses $m_i$ and spins $J_i$ one finds \cite{JHEPCC1}:
\begin{eqnarray}
\frac{d\CC}{d\ln\mu}&& =\frac{1}{(4\pi
)^{2}}\,\sum_{i}\,A_{i}\,m_{i}^{4}
 \label{newRGUV}
\end{eqnarray}
where
\begin{equation}\label{Ai}
A_i=(-1)^{2J_i}(J_i+1/2)\,n_{J_i}\,N_c\,,
\end{equation}
with $n_{\{0,1,1/2\}}=(1,1,2)$ and $N_c=1,3$ for uncolored and
colored particles respectively. However, this equation applies
only if the masses $m_i$ do correspond to active degrees of
freedom (\textit{d.o.f.}), namely if the RG scale $\mu$ satisfies
$\mu\gg m_i$ for all $i=1,2,3,...$. Any mass that does not
satisfy this condition must be dropped from that RGE.  At the
same time this raises two issues: i) One is the meaning of the RG
scale $\mu$ in cosmology, and the other is: ii) what happens with
the decoupled contributions, i.e. those heavy masses that satisfy
$M_i>\mu$ for a given value of the RG scale.  This question has
been addressed in Ref.\,\cite{JHEPCC1,Babic} where it was first
recognized that the contribution from the heavy degrees of
freedom could be relevant, especially in the context of the
cosmological constant problem, where every ``small'' contribution
(for the standards of high energy physics) can be very big (for
the standards of cosmology). This reflection suggests that
Eq.\,(\ref{newRGUV}) can only be a ``sharp cutoff'' approximation
to the complete RGE for the CC. In other words, it is not enough
to (abruptly) couple or decouple the various degrees of freedom
following a $\theta$-function procedure like in the
renormalization group equations that we are used to in the
Minimal Subtraction ($\,\overline{MS}\,$)-scheme\,\cite{Collins}.
Here the smoothing of the various contributions could play a
role. In fact, this smoothing is perfectly possible if one adopts
a more physical renormalization scheme, like the Momentum
Subtraction Scheme\,\cite{Manohar}. It can be of course much more
cumbersome, but it provides a well-defined prescription to
``connect'' and ``disconnect'' the various \textit{d.o.f.}
automatically, i.e. not just by hand. Then, from general
arguments of effective field theory, in combination with the
Appelquist-Carazzone (AC) decoupling theorem\,\cite{AC}), we
expect that the generalization of the RGE for any energy regime
is the following:
\begin{eqnarray}
\frac{d\CC}{d\ln\mu}&& =\frac{1}{(4\pi
)^{2}}\left(\sum_{i}\,A_{i}\,m_{i}^{4}
+\mu^{2}\sum_{j}%
\,B_{j}M_{j}^{2}\,\,+\mu^{4}\sum_{j}
\,C_{j}+\mu^{6}\sum_{j}\frac{\,D_{j}}{M_{j}^{2}}\,\,+...\right)
\label{newRG1}
\end{eqnarray}
where the sums are taken over all massive fields;
$\,A,B,C,D,...\,$ are constant coefficients, and  $\mu$ is the
energy scale associated to the RG running. The \textit{r.h.s.} of
(\ref{newRG1}) defines the $\bCC$-function for $\CC$, which is a
function of the masses and in general also of the ratios of the
RG scale and the masses, $\mu/M_i$.

Concerning the interpretation of the RG scale $\mu$ in cosmology,
within this, more physical, RGE we adopt the proposal of
Ref.\,\cite{JHEPCC1}, namely we identify $\mu$ with the value of
the Hubble parameter at the corresponding epoch:
\begin{equation}
\mu \sim H(t)  \label{muH}\,.
\end{equation}
From Eq.\,(\ref{totalR}) and (\ref{FL1}) it is clear, within
order of magnitude, that in the FLRW cosmologies this is a
particularization of identifying $\mu$ with $R^{1/2}$, the latter
being a perfectly sensible (and invariant) way to define the RG
scale in a general cosmological framework\,\footnote{ Scale
(\ref{muH}) has also been used successfully in other frameworks,
e.g. in \cite{shocom} to describe the decoupling of massive
particles in anomaly-induced inflation --or modified Starobinsky
model\,\cite{star}.}. On physical grounds, what this implies is
to identify $\mu$ in cosmology with the typical energy-momentum of
the cosmological gravitons for a given metric, in this case the
FLRW one (\ref{FLRWm}). With this Ansatz, it is clear that only
particles whose masses are below the value of the Hubble parameter
($m_i<H$) can be active \textit{d.o.f.} at the given epoch.
Notice that this makes extremely difficult to find out active
particles, from the point of view of the RG, in any cosmological
epoch!  For example, the present value of the Hubble parameter is
$\,H_0\sim 10^{-33}\,eV$. The latter is 30 orders of magnitude
smaller than the mass of the lightest neutrino, 41 orders of
magnitude smaller than the QCD scale and 61 orders of magnitude
smaller than the Planck scale. Obviously, all massive particles
decouple the same way!
 Nevertheless, this effective decoupling of all the \textit{d.o.f.}
corresponding to the SM particles (in fact of all particles well
below the Planck scale!) does not necessary mean that there is no
running of the CC. The general RGE\,(\ref{newRG1}) contains
indeed the clue to the possible solution of this problem.
Namely,  $\CC$ could effectively run mainly due to the heaviest
\textit{d.o.f.}, represented by the remaining terms on the
\textit{r.h.s.} of (\ref{newRG1}) other than those in
(\ref{newRGUV}).  Looking at the chain of decoupled terms, we
immediately discover the dominant ones in the $\bCC$-function,
\begin{equation}
  \bCC =\frac{1}{(4\pi)^{2}}
\left(\mu^{2}\sum_{j}%
\,B_{j}M_{j}^{2}+...\right) \label{newRG2}
\end{equation}
which we call the ``soft decoupling'' contributions to the RGE of
the CC. These are rather peculiar, as they grow quadratically
with the masses of the various particles. Notice that the
existence of these soft decoupling contributions stems from the
dimension-four nature of the CC. There is no other parameter
either in the SM or in the GUT models with such property.

The structure of the RGE\ (\ref{newRG1}) is dictated by the the
AC-decoupling theorem \cite{AC,Collins} and general covariance.
Indeed, dimensional analysis is not enough to explain the most
general structure of $\bCC$. The fact that only even powers of
$\mu$ are involved stems from the covariance of the effective
action under the identification (\ref{muH}). Odd powers of $\mu=H$
cannot appear after integrating out the higher derivative terms
as they must appear bilinearly in the contractions with the
metric tensor. In particular, covariance forbids the terms of
first order in $H$. As a result the expansion must start at the
$H^2$-order. In short, when applying the AC theorem in its very
standard form to the computation of the $\bCC$-function of the
cosmological constant, the decoupling of the heavy masses does
still introduce inverse power suppression, but since the
$\bCC$-function itself is proportional to the fourth power of
these masses it eventually entails a decoupling law
$1/M^{2n-4}_j$, and so the $n=1$ and $n=2$ terms do not decouple
in the ordinary sense whereas the $n=3$ and above terms do. The
upshot is that the $n=1$ (soft decoupling terms) dominate the
$\bCC$-function.

What about the phenomenological significance of this RG framework
for the cosmological constant? The remarkable thing about the soft
decoupling terms is that if we assume that there is physics near
the Planck scale, characterized by a set of (fermion and boson)
fields with masses near the Planck scale ($M_i\lesssim M_P$),
then we can realize the numerical ``coincidence''
\begin{equation}\label{numeric}
c\,H_0^{2}M_i^{2}/(4\pi)^2\simeq c\,\left(1.5\
10^{-42}\,GeV\,\times 1.2\
10^{19}\,GeV\right)^2/\left(4\pi\right)^2\sim 10^{-47}\,GeV^4\sim
\CC_0
\end{equation}
for some $c={\cal O}(1-10)$. Put another way: using the RG scale
Ansatz (\ref{muH}), the contribution from (near-) Planckian size
masses to the soft decoupling terms is just of the order of the
present day value of the cosmological constant -- Cf.
Eq.\,(\ref{CCvalue}). This means that, thanks to the soft
decoupling terms (\ref{newRG2}), there is the possibility to
obtain an effective running of the CC which changes its value by
an amount that is of the order of the CC value itself. Hence a
smooth running of the CC is feasible without requiring any fine
tuning of the parameters.

Let us introduce the following mass parameter:
\begin{equation}\label{Mdef}
M\equiv\sqrt{\left|\sum_i\,c_i\,M_i^2\right|}\,.
\end{equation}
The mass $M_i$ of each superheavy particle may be smaller than
$M_P$ and the equality, or even the effective value $\,M\gtrsim
M_P$, can be achieved due to the multiplicities of these
particles. At the end of the day we expect from (\ref{newRG2})
that the RGE of the CC is dominated by the term
\begin{equation}\label{RG01}
\frac{d\CC}{d{\ln H}}= \frac{1}{(4\pi)^{2}}\
\sigma\,H^{2}M^{2}+...=\frac{3\,\nu}{4\,\pi}\,H^{2}\,M_P^2+...
\end{equation}
where the signature $\,\sigma=\pm 1$ of the $\bCC$-function
depends on whether the fermions ($\sigma=-1$) or bosons
($\sigma=+1$) dominate at the highest energies. In the previous
RGE we have defined the fundamental (dimensionless) parameter
\begin{equation}\label{nu}
\nu\equiv \frac{\sigma}{12\,\pi}\,\frac{M^2}{M_P^2}\,,
\end{equation}
which, as we shall see, traces out the presence of the RG effects
on the modified cosmology as well as its departure from the
standard FLRW framework.

A few reflections are now in order. The physical interpretation
of the renormalization group in curved space-time requires the
formalism beyond the limits of the well-known standard
techniques. These techniques are essentially based on the minimal
subtraction $\,\overline{MS}\,$ scheme of renormalization
\cite{Collins,book}, which does not admit to observe the
decoupling. The physical interpretation of the renormalization
group in the higher derivative sector of the vacuum action
(\ref{Svac}) can be achieved through the calculation of the
polarization operator of gravitons arising from the matter loops
in linearized gravity \cite{apco}. In this case one can perform
the calculations in the physical mass-dependent renormalization
scheme (specifically, in the momentum subtraction scheme, in
which $\mu$ is traded for an Euclidean momentum $p$), obtain
explicit expressions for the physical beta-functions and observe
the decoupling of massive particles at low energies. In the
mass-dependent scheme one has direct control of the functional
dependence of the $\,\beta$-functions on the masses of all the
fields, i.e. one knows explicitly the functions
$\,\beta_k(m_i;p^2/M^2_j)\,$ for the higher derivative terms
$a_k$ in (\ref{Svac}). Quite reassuring is the fact that these
momentum subtraction $\,\,\beta$-functions boil down, in the UV
limit, to the corresponding $\,\,\beta$-functions in the MS
scheme \cite{apco}.  At present there is no method of calculations
on the non-flat background compatible with the physical
renormalization scheme. In this situation our phenomenological
approach looks rather justified. The covariance of the effective
action forbids the first order in $\,H\,$ corrections. Then,
recalling that the UV contribution to $\,\beta_\Lambda\,$ from a
particle of mass $\,m\,$ is $\beta_\Lambda\sim m^4$, the
low-energy contributions to $\,\beta_\Lambda\,$ must be supressed
by, at least, the factor
\begin{equation}
\square/m^2\,\,\sim\,\,H^2/m^2\,, \label{endequation}
\end{equation}
and hence the overall low-energy $\,\beta_\Lambda\,$ acquires the
form $H^2\,m^2$, Eq.\,(\ref{RG01}). The same form of decoupling
can be expected for the induced counterpart $\CC_{ind}$ of the
CC. The renormalization group equations for $\CC_{ind}$ and
$\CC_{vac}$ in (\ref{lambdaphys}) are independent, and the
relation emerges only at the moment when we choose the initial
point of the RG trajectory for the vacuum counterpart
\cite{JHEPCC1,cosm}. In fact, the Feynman diagrams corresponding
to the $\CC_{ind}$ are, at the cosmic scale, very similar to the
vacuum ones, because the Higgs field $\phi$ is in the vacuum
state, $<\phi>\neq 0$, and the non-zero contribution of these
diagrams is exclusively due to the momenta coming from the
graviton external lines. Finally, there is no real need to
distinguish induced and vacuum CC's in the present context. In
both cases the absence of the $n=0$ order (non-supressed)
contributions in (\ref{newRG1}) is required by the apparent
correctness of the Einstein equations and the smallness of the
observable CC.

Let us finally remark on the running in other sectors. In the
basic equation of energy conservation (\ref{Bronstein}) one could
have enforced the conservation of matter by itself, but if one
still wishes an evolving CC this could only be posible at the
expense of a time-varying Newton's constant, which could give
rise to various interesting effects\,\cite{Reuter03a,Stefancic2}.
However, in our framework this is not granted. For the inverse
Newton constant $1/G\,$ the running is irrelevant because
$1/G\sim M_P^2$ is very large and the effect of the running is
relatively small {as it was demonstrated in \cite{JHEPCC1}}. On
the other hand, the relevance of the higher derivative terms at
low energies is supposed to be negligible, as it was recently
shown in \cite{Pelinson} for the perturbations of the conformal
factor. In contrast, the soft decoupling for the CC does matter
because the CC is very small and any running, even the very small
one that we get, is sufficient to produce some measurable effect
at large redshifts. For precisely this reason we have to use the
Friedmann equation, in combination with the full conservation law
including the variable cosmological term, and no other terms.

\section{A ``renormalized'' FLRW cosmology with running $\CC$}

After introducing and discussing our theoretical framework, let us
now solve explicitly for the cosmological evolution equations for
the matter and vacuum energy densities ensuing from our
semiclassical FLRW model\,\cite{Letter,CCfit}. Let us first of
all assume that we are in the matter era, then $p=0$. We can now
solve for the simultaneous system formed by equations
(\ref{FL1}), (\ref{Bronstein}) and our basic RGE (\ref{RG01}). It
is convenient to eliminate the time variable in favor of the
redshift $z$, using $a_0/a=1+z$.  The result is obtained after
straightforward calculation. For the matter density we find
\begin{equation}
\rho(z;\nu) \,=\,\Big(\rM^0+\frac{\kappa}{1-3\nu}\,\rho_c^0\Big)
\,(1+z)^{3(1-\nu)} -\frac{\kappa}{1-3\nu}\,\rc^0\,(1+z)^2\,,
\label{rhoznu}
\end{equation}
where we have introduced the parameter
\begin{equation}\label{bdef}
\kappa \equiv -2\,\nu\OK^0\,.
\end{equation}
It is nonzero only for spatially curved universes, and it depends
on the fundamental parameter $\nu$, Eq.\,(\ref{nu}). The
arbitrary constant in (\ref{rhoznu}) has been determined by
imposing the condition that at $z=0$ we must have $\rho=\rM^0$.
Similarly, solving for the $\nu$-dependent $\CC$ as a function of
the reshift, with the initial condition $\CC(z=0)=\CC_0$, we find
\begin{equation}
\CC(z;\nu)= \CC_0+\rM^0\,f(z)\,+\rc^0\,g(z)\,, \label{Lambdaznu}
\end{equation}
with
\begin{eqnarray}
f(z)=\frac{\nu}{1-\nu}\,\left[\left(1+z\right)^{3(1-\nu)}-1\right]\,,
\label{f}
\end{eqnarray}
\begin{eqnarray}
g(z)\,\,=-\frac{\kappa}{1-3\nu}\,\left\{\frac{z\,(z+2)}{2}\,
+\,\frac{\nu}{1-\nu}\,\left[\left(1+z\right)^{3(1-\nu)}-1\right]
\right\}\,. \label{g}
\end{eqnarray}
Furthermore, since we have already obtained the functions
$\rho(z;\nu)$ and $\CC=\CC(z;\nu)$, we can use them in (\ref{FL1})
to get the explicit $\nu$-dependent Hubble function $H(z;\nu)$. It
reads
\begin{eqnarray}
\label{Hzzz} H^2(z;\nu)&=& H_0^2\,\left\{1+\Omega_M^0\,
\frac{\left(1+z\right)^{3\,(1-\nu)}-1}{1-\nu}  \right.
\nonumber\\
&&\left. +\frac{1-\Omega_M^0-\Omega_{\Lambda}^0}{1-3\,\nu}
\left[(1+z)^2-1-2\nu
\,\frac{\left(1+z\right)^{3\,(1-\nu)}-1}{1-\nu}\right] \right\}\,.
\end{eqnarray}
For flat universes (\ref{Hzzz}) shrinks to
\begin{eqnarray}
\label{Hzzzflat} H^2(z;\nu)&=& H_0^2\,\left\{1+\Omega_M^0\,
\frac{\left(1+z\right)^{3\,(1-\nu)}-1}{1-\nu}  \right\} \,,
\end{eqnarray}
whereas the standard result in this case is\,\cite{Peebles}
\begin{eqnarray}\label{HzSS}
H^2(z)&=& H^2_0\,\left\{1+\Omega_M^0\,\left[(1+z)^3-1\right]
\right\}\,,
\end{eqnarray}
which is indeed recovered from Eq.\,(\ref{Hzzzflat}) in the limit
$\nu=0$.

Notice that the kind of general behaviour that we have found for
the CC evolution is of the form:
\begin{equation}\label{CCbehave}
\CC (t)=\CC_0+\xi\,[\,H^2(t)-H^2(t_0)]\,\,M_P^2\,,
\end{equation}
where the coefficient $\xi$ is proportional to $\nu$:
\begin{equation}\label{xi}
\xi=\frac{3\,\nu}{8\pi}\,.
\end{equation}
Let us point out that many similar phenomenological equations for
the CC evolution have been tried in the
literature\,\cite{CCphenom,Overduin}, but the interesting thing
in our particular realization is that it is motivated by the
Renormalization Group, and therefore the structure
(\ref{CCbehave}) may emerge from a fundamental QFT formulation.

Before applying our model to test the status of the present
universe, it is important to check whether the model can be made
compatible with nucleosynthesis. In fact, a non-vanishing $\nu$
may have an impact not only in the matter-dominated (MD) era, but
also in the radiation-dominated (RD) epoch. To this end we recall
that in the RD era Eq.\,(\ref{Bronstein}) must include the $p\neq
0$ term. For photons the radiation density $\rho_R$ is related to
pressure through $p=(1/3)\,\rho_R$. With this modification we may
solve again the differential equations in the radiation era. In
this case it is more natural to express the above result in terms
of the temperature. For the radiation density we find
\begin{eqnarray}\label{rhozR2}
\rho_R(T;\nu) = \frac{\pi ^{2}}{30}\,g_{\ast
}\,T^{4}\,\left(\frac{T_0}{T}\right)^{4\,\nu}
+\frac{\kappa}{2-4\nu}\,\left[\left(\frac{T}{T_0}\right)^{4\,(1-\nu)}
-\left(\frac{T}{T_0}\right)^2\right]\,,
\end{eqnarray}
where $T_{0}\simeq 2.75\,K=2.37\times 10^{-4}\,eV$ is the present
CMB temperature.  Of course for $\nu\rightarrow 0$ we recover the
standard result
\begin{equation}\label{rhozphot}
\rho_R(T)=\rho_R^0\,\left(\frac{T}{T_0}\right)^{4}=\frac{\pi
^{2}}{30}\,g_{\ast }\,T^{4}\,,
\end{equation}
with $\,g_{\ast }=2\,$ for photons and $\,g_{\ast }=3.36\,$ if we
take neutrinos into account. For the CC,
\begin{equation}\label{LambdazR1}
\Lambda_R(T;\nu)=\Lambda_0+\rM^0\,f_R(T)+\rc^0\,g_R(T)\,,
\end{equation}
with
\begin{equation}\label{fzR}
 f_R(T)=\frac{\nu}{1-\nu}
\,\left[\left(\frac{T}{T_0}\right)^{4\,(1-\nu)}-1\right]\,
\end{equation}
and
\begin{equation}
\label{gzR} g_R(T)\,=\,-\,
\frac{\kappa}{2-4\nu}\,\left\{\frac{T^2-T_0^2}{T_0^2}
\,-\,\frac{\nu}{1-\nu}
\,\left[\left(\frac{T}{T_0}\right)^{4\,(1-\nu)}
-1\right]\right\}\,.
\end{equation}
When comparing the relative size of the CC, Eq.
(\ref{LambdazR1}), versus the radiation density, Eq.
(\ref{rhozR2}), at the time of the nucleosynthesis we naturally
require that the former is smaller than the latter. Therefore,
for small $\kappa$, we impose
\begin{equation}\label{ratioCCrho2}
\left|\frac{\Lambda_R(T)}{\rho_R(T)}\right|\simeq
\left|{\nu}\right|\ll 1\,.
\end{equation}

\section{Numerical results}
\label{sect:numerical}

After the nucleosynthesis restriction on the $\nu$-parameter, the
question is whether there is still some room for useful
phenomenological considerations at the present matter epoch.
Fortunately, the answer is yes. A convenient fiducial value for
the cosmological index $\nu$ is:
\begin{equation}\label{nuzero}
\nu_0\equiv\frac{1}{12\,\pi}\simeq  2.6\times 10^{-2}\,,
\end{equation}
which corresponds to $M=M_P$ in (\ref{nu}). We will use this
value, and some multiples of it (with both signs) to perform the
numerical analysis\,\cite{CCfit}. The maximum value that we will
tolerate for $\nu$ is $\nu=0.1$, which still respects the
nucleosynthesis bound (\ref{ratioCCrho2}). Let us also
circumscribe the numerical analysis to the flat case, $k=0$. The
evolution of the matter density and of the CC is shown in
Fig.\,\ref{fig:evmupla1}a,b. These graphics illustrate
Eq.\,(\ref{rhoznu}) and (\ref{Lambdaznu}). As a result of
allowing a non-vanishing $\bCC$-function for the CC
(equivalently, $\nu\neq 0$) there is a simultaneous, correlated
variation of the CC and of the matter density.

Comparing with the standard model case $\,\nu=0\,$ (see the
Fig.\,\ref{fig:evmupla1}a,b), we see that for a negative
cosmological index $\,\nu\,$ the matter density grows faster
towards the past ($z\rightarrow\infty$) while for a positive
value of $\,\nu\,$ the growing is slower than the usual
$\,(1+z)^3$. Looking towards the future ($z\rightarrow -1$), the
distinction is not appreciable because for all $\nu$ the matter
density goes to zero. The opposite result is found for the CC,
since then it is for positive $\nu$ that $\CC (z;\nu)$ grows in
the past, whereas in the future it has a different behaviour,
tending to different (finite) values in the cases $\,\nu<0\,$ and
$\,0<\nu<1$, while it becomes $\,\,-\infty\,$ for $\,\nu\geq 1$
(not shown).

\FIGURE[h]{
\mbox{\resizebox*{0.5\textwidth}{!}{\includegraphics{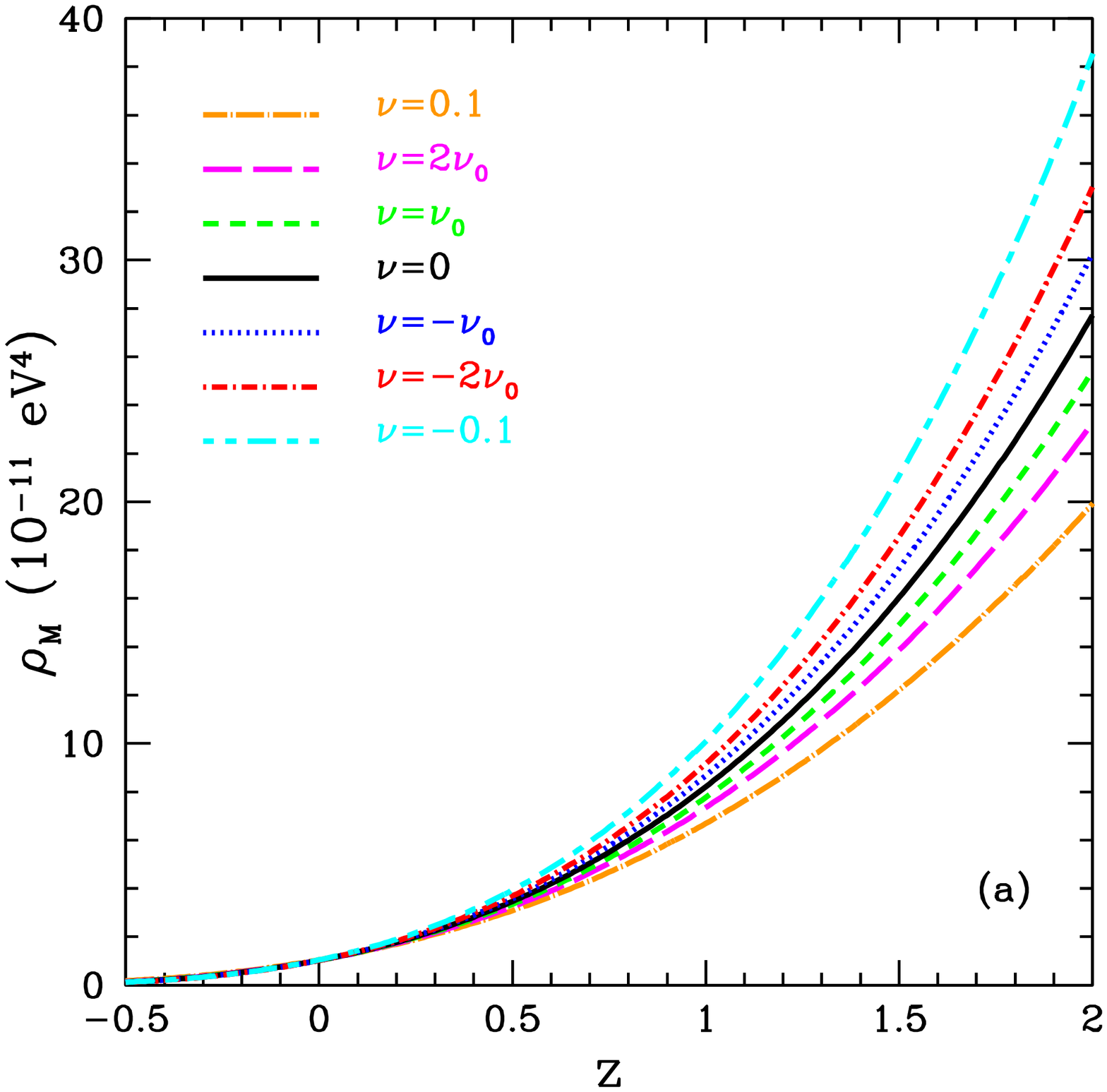}}
   \resizebox*{0.5\textwidth}{!}{\includegraphics{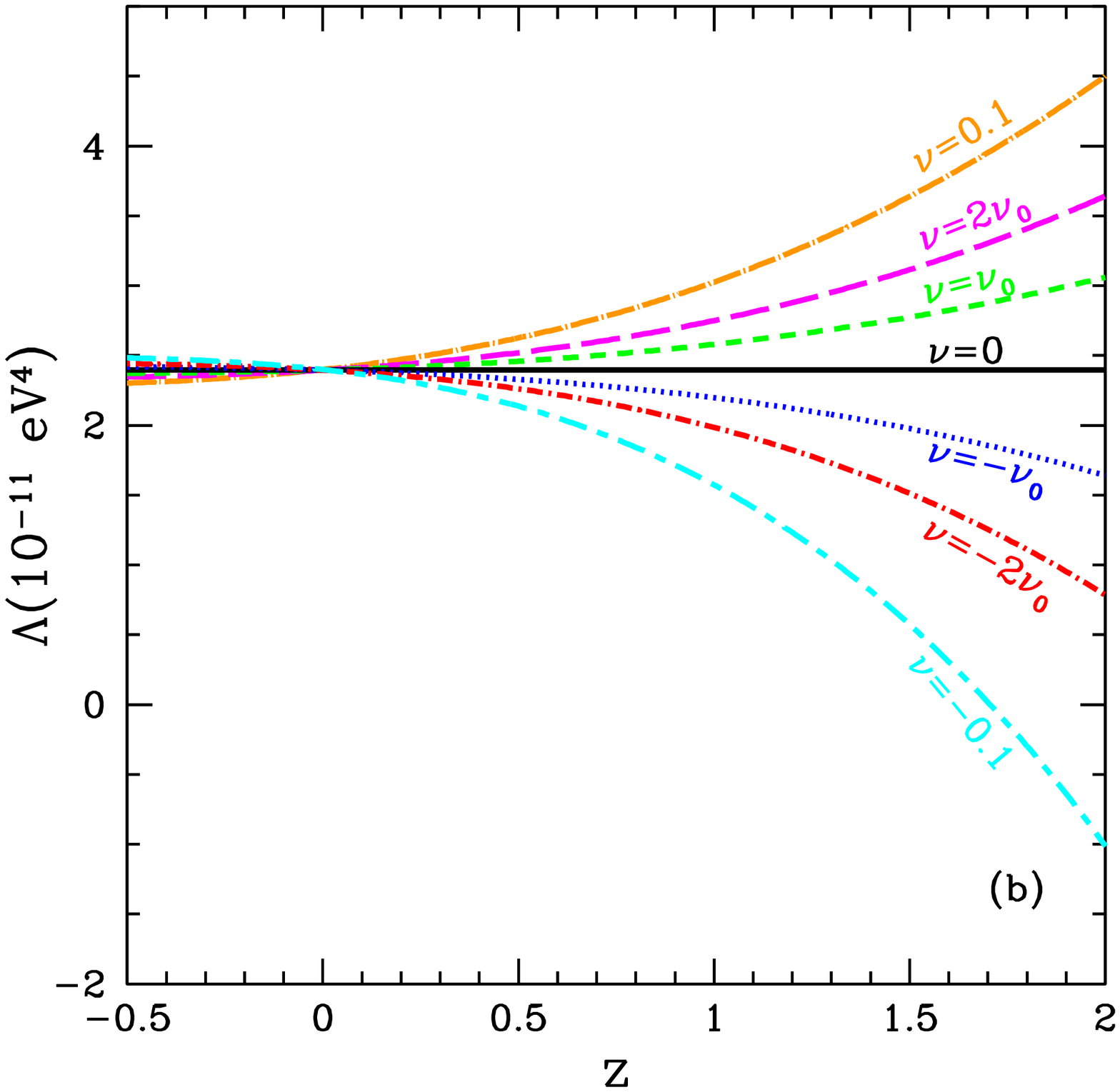}}}
\caption{(a),(b) Future and past evolution of the matter density
$\rM (z;\nu)$ and the cosmological constant $\CC (z;\nu)$ for a
flat universe ($k=0$) and for different values of the fundamental
parameter $\nu$ of our model ($\nu_0$ is defined in
Eq.(\ref{nuzero})). In both cases $\Omega_M^0=0.3$ and
$\Omega_{\Lambda}^0=0.7$. The solid line represents the standard
model case ($\CC={\rm const.}$), whereas the various kinds of
dots/dashes represent different amounts of evolution of
$\CC$\,\cite{CCfit}.} \label{fig:evmupla1}}

\FIGURE[h]{ \centerline{
\mbox{\resizebox*{0.5\textwidth}{!}{\includegraphics{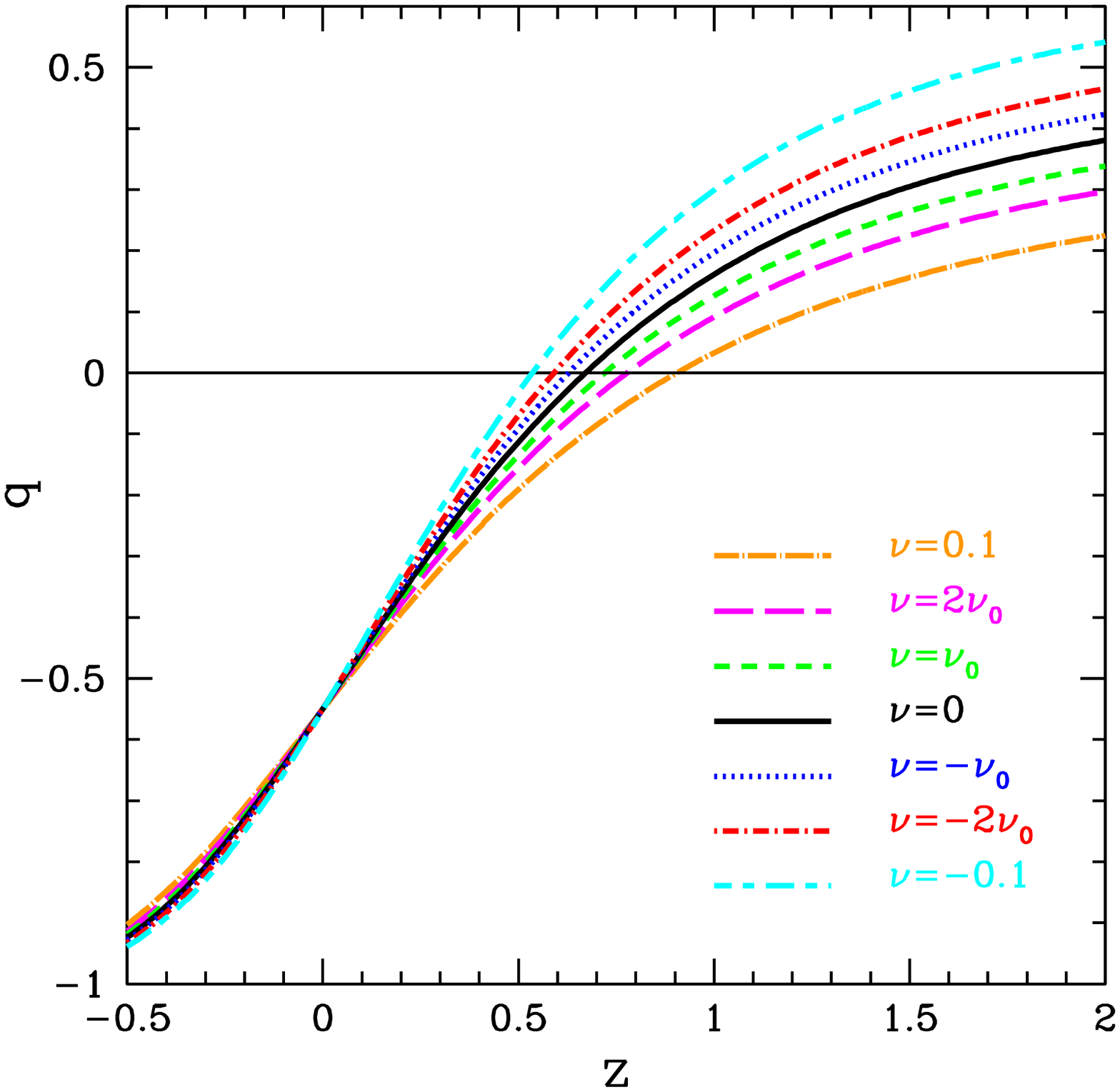}}}
} \caption{Evolution of the deceleration parameter, $q$, for a
flat universe ($k=0$) with $\Omega_M^0=0.3$ and
$\Omega_{\Lambda}^0=0.7$. We see that the transition from a
decelerated ($q>0$) to an accelerated ($q<0$) universe takes place
earlier in time (larger redshifts) for larger $\nu>0$ (see text
for the exact values)\,\cite{CCfit}. \label{fig:qHpla}}}

In the phenomenologically most interesting case $\,|\nu|<1\,$ we
always have a null density of matter and a finite (positive) CC
in the long term future, while for the far past yields
$\,\CC=\pm\infty\,$ depending on the sign of $\,\nu$.  In all
these situations the matter density safely tends to $\,+\infty$.
One may worry whether having infinitely large CC and matter
density in the past may pose a problem to structure formation.
From Fig.\,\ref{fig:evmupla1}a,b it is clear that there should
not be a problem at all since in our model the CC remains always
smaller than the matter density in the far past, and in the
radiation epoch $z>1000$ we reach the safe limit
(\ref{ratioCCrho2}). Actually the time where $\CC (z;\nu)$ and
$\rM (z;\nu)$ become similar is very recent.

We may consider another relevant exponent describing how the
universe evolves, the deceleration parameter {\it
q}\,\cite{Peebles}. This one is fully sensitive to the kind of
high-z SNe Ia data under consideration. By definition,
\begin{equation}
\label{pa1}
q(z;\nu)=-\frac{\ddot{a}}{a\,H^2(z;\nu)}=\,-1-\frac{\dot{H}}{H^2}
\,=\,-1+\frac12\,(1+z)\,\frac{1}{H^2(z;\nu)}\,\frac{dH^2(z;\nu)}{dz}\,,
\end{equation}
where the $\nu$-dependent Hubble parameter $H^2(z;\nu)$ is given
in Eq.\,(\ref{Hzzz}). It is interesting to look at the deviations
of the deceleration parameter with respect to the standard model.
It is well-known that there are already some data on Type Ia
supernovae located very near the critical redshift $z^{*}$ where
the universe changed from deceleration to acceleration
\cite{transition}. But of course the precise location of $z^{*}$
depends on the FLRW model and variations thereof. In our case
$\,z^*\,$ should depend on our cosmological index $\nu$, i.e.
$z^{*}=z^{*}(\nu)$. For simplicity in the presentation, let us
consider the flat case. The transition point between accelerated
and decelerated expansion is a function of $\,\nu\,$: the more
negative is $\,\nu$, the more delayed is the transition (closer
to  our time)-- see Fig.\,\ref{fig:qHpla}. If $\,\nu>0$, the
transition occurs earlier (i.e. at larger $z$). While in the
standard case, and for a flat universe, the transition takes
place at redshift
\begin{equation}\label{zstar}
z^{*}=-1+\sqrt[3]{2\,\frac{\OL^0}{\OM^0}}\simeq 0.67\,,
\end{equation}
it would have occurred at $z=0.72$ and $z=0.78$ for $\nu=\nu_0$
and $\nu=2\nu_0$ respectively, and at $z=0.63$ and $z=0.59$ for
$\nu=-\nu_0$ and $\nu=-2\nu_0$ (Cf. Fig.\,\ref{fig:qHpla}). For
$\nu=(-0.1,+0.1)$ the effect is quite large, namely the
transition would be at $z=(0.53,0.91)$ and hence there is a
correction of $(-21\%,+36\%)$ with respect to the standard case.

 \vspace{0.5cm}

\TABULAR[h]{l|c|c|c|c} { \hline \hline
              &      &           & \\
 Distribution~~        & Data      & Prior $\sigma_{\Omega_M}$ &
        $\nu$ & $\sigma_{\nu}$  \\
              &      &           & \\
\hline
               &                   &      &           &             \\
{\bf SNAP}     & 50 SNe $0<z<0.2$  &      &           &             \\
(1 year)      &1800 SNe $0.2<z<1.2$&      &           &             \\
               &50 SNe $1.2<z<1.4$ &      &           &             \\
               &15 SNe $1.4<z<1.7$ & None & $0.1$ &~~$\pm0.10$  \\
               &                   &      &           &             \\
{\bf SNAP}     & as above          & 0.03 & $0.1$ &~~$\pm0.06$  \\
               &                   &      &           &             \\
{\bf SNAP}     &  3 years          & None & $0.1$ &~~$\pm0.06$  \\
               &                   &      &           &             \\
{\bf SNAP}     &  3 years          & 0.03 & $0.1$ &~~$\pm0.04$  \\
               &                   &      &           &             \\
{\bf Distr.1}  & 50 SNe $0<z<0.2$  &      &           &             \\
             &2000 SNe $0.2<z<1.7$ & 0.03 & $0.1$ &~~$\pm0.05$  \\
               &                   &      &           &             \\
{\bf Distr.2}  & 250 SNe $0<z<1$   &      &           &             \\
               &1750 SNe $1<z<2$   & 0.03 & $0.1$ &~~$\pm0.02$  \\
               &                   &      &           &             \\
\hline \hline
 }
{ Determination of $\nu$ with SNAP data and with other two
distributions. In all cases we assume a flat universe. When a
prior on $\Omega_M$ and its error is added we use $\Omega_M=0.3\pm
0.03$\,\ \cite{CCfit}.} \label{tab:resultats}

It should be clear that our approach based on a variable CC
departs from all kind of quintessence-like approaches, in which
some slow--rolling scalar field $\chi$ substitutes for the CC. In
these models, the dark energy is tied to the dynamics of the
self-conserved $\chi$ field;  i.e. in contrast to
(\ref{Bronstein}) there is no transfer between $\chi$-dark energy
and ordinary forms of energy. The phenomenological equation of
state is defined by $p_{\chi}=w_{\chi}\,\rho_{\chi}$. The term
$-2\,\CC$ on the \textit{r.h.s.} of Eq.\,(\ref{newforce3}) must
be replaced by
$\rho_{\chi}+3\,{p}_{\chi}=(1+3\,w_{\chi})\rho_{\chi}$. In order
to get accelerated expansion in an epoch characterized by $p=0$
and $\rho\rightarrow 0$ in the future, one must require
$-w_{-}\leq w_{\chi}\leq -1/3$, where usually $w_{-}\geq -1$ in
order to have a canonical kinetic term for $\chi$\,\footnote{One
cannot completely exclude ``phantom matter-energy'' ($w_{-}< -1$)
and generalizations thereof\,\cite{Stefan}.}. The particular case
$w_{\chi}=-1$ corresponds to a quintessence field exactly
mimicking the cosmological constant term. In fact, for a
cosmological term $\CC$, whether constant or variable, the only
possible equation of state is the one corresponding to
$w_{\CC}=-1$, as it obvious from the definition of
$\tilde{T}_{\mu\nu}$ in Eq.\,(\ref{EE}). Although $p_{\chi}$ and
$\rho_{\chi}$ are related to the energy-momentum tensor of
$\chi$, the dynamics of this field is unknown because the
quintessence models do not have an explanation for the value of
the CC. Therefore, the barotropic index $w_{\chi}$ is not known
from first principles. In particular, one cannot exclude it may
have a redshift dependence, which can be parametrized with two
parameters as follows:
\begin{equation}\label{wexpansion}
\frac{p_{\chi}}{\rho_{\chi}}\equiv w_{\chi}=w_0+w_1 (1-a)=w_0+w_1
\frac{z}{1+z}\,.
\end{equation}
Finding a non-vanishing value of $w_1$ implies a redshift
evolution of the equation of state for the $\chi$ field
\,\cite{WMAP}. For completeness, consider the modification on the
Hubble parameter introduced by quintessence models. In this case,
and in the flat case, it is easy to see that
\begin{eqnarray}
\label{Hzzzquint} H^2(z;w_0,w_1)&=&
H_0^2\,\left\{\Omega_M^0\,(1+z)^3\right.\nonumber\\
&&+\left.(1-\Omega_M^0)(1+z)^{3(1+w_0+w_1)}\,\exp\left[-3\,w_1\,\frac{z}{1+z}\right]
\right\} \,.
\end{eqnarray}
This equation reduces to the standard one (\ref{HzSS}) for
$w_{\chi}=-1$ ($w_0=-1,w_1=0$), as expected. Comparison between
(\ref{Hzzzquint}) and (\ref{Hzzzflat}) can be useful to identify
the differences between RG models and quintessence models of the
dark energy.

Fitting cosmological models to high-z supernovae data is usually
performed via the so-called magnitude-redshift
relation\,\cite{p99,Riess98}. One starts from the notion of
luminosity distance, $d_L$\, related to the received flux
${\cal{F}}$ and the absolute (intrinsic) luminosity ${\cal{L}}$
through the geometric definition \cite{Peebles}:
\begin{equation}\label{flux}
{\cal{F}}=\frac{{\cal{L}}}{4\pi d_L^2}\,.
\end{equation}
Then the logarithmic relation between flux and the (theoretical)
apparent magnitude reads
\begin{equation}\label{magnitud}
m^{th}(z,H_0,\OM^0,\OL^0) \,=\, {\cal{M}}\,+\,
 5\log_{10}\left[H_0\, d_L(z,H_0,\OM^0,\OL^0)\right]\,,
\end{equation}
where $M$ is the absolute magnitude (believed to be constant for
all Type Ia supernovae, assuming they are real ``standard
candles''), and
\begin{equation}\label{m0}
{\cal{M}} = M - 5\log_{10} H_0 \,+\, 25=M - 5\log_{10} h_0 \,+\,
42.38\,.
\end{equation}
The model dependence is encoded in the luminosity-distance
function $d_L=d_L(z,H_0,\OM^0,\OL^0)$, given in our case
by\,\cite{CCfit}
\begin{equation}\label{dist}
d_L(z,H_0,\OM^0,\OL^0;\nu) = \frac{1+z}{H_0 \sqrt{|\OK^0|}}\ \Psi
\left(
 \sqrt{|\OK^0|} \int_{0}^{z}{\frac{H_0 ~ dz'}{H(z',\OM^0,\OL^0;\nu)}}
 \right)\,,
\end{equation}
with
\begin{equation}\label{kcases}
\Psi(x)=\left\{\begin{array}{lr}
    \sin{x},~~~\OK^0<0\\
     x,~~~~~~~~\OK^0=0\\
    \sinh{x},~~\OK^0>0\,.
       \end{array}\right.
\end{equation}
Here the expansion rate $H(z,\OM^0,\OL^0;\nu)$ is given by Eq.
(\ref{Hzzz}). In the flat case it reduces to the two-parameter
function (\ref{Hzzzflat}), and in quintessence models to the
three-parameter function (\ref{Hzzzquint}). Usually a prior on
$\OM^0$ (from cluster dynamics) can be accepted (e.g.
$\OM^0=0.3$), which narrows down the number of parameters to one
(RG) and two (quintessence).

We can use the magnitude-redshift relation defined above to test
various distributions, including the one foreseen by
SNAP\,\cite{SNAP}. In order to determine the cosmological
parameters we use a $\chi^2$-statistic test, where $\chi^2$ is
defined by the difference between the theoretical apparent
magnitude and the observed one (see Ref.\,\cite{CCfit} for
details). The idea is that the difference between models grows
with redshift. The existing sample of SN Ia data is amply
compatible with $\nu\neq 0$, but it does not pin down a narrow
interval of values for this parameter. Trying, however, over
several large (simulated) distributions of supernovae, and
marginalizing over ${\cal M}$, one obtains the results shown in
Table \ref{tab:resultats}. Distribution 1 is very similar to the
SNAP one but with most of the data homogeneously distributed
between $z=0.2$ and $z=1.7$. Distribution 2 extends data up to
redshift $z=2$. We see that we can determine $\nu$ to within $\pm
(20-60)\%$ for $\nu=0.1$, depending on the distribution. (Smaller
values of $\nu$ imply smaller precision.)  The situation is
similar to the determination of the evolution of the equation of
state, Eq.\,(\ref{wexpansion}). Indeed, if one performs a general
(model-independent) fit of the present SN Ia data to quintessence
models, leaving free the two parameters $w_0$ and $w_1$ in Eq.
(\ref{wexpansion}), one finds that the values $w_0>-1/3$
(decelerated universe) are ruled out at a high significance level
(for $\OM^0<0.4$), thereby supporting the existence of dark
energy. Nevertheless, the very same fit is highly insensitive to
$w_1$\,\cite{Freeman,Padmanabhan}.  On the other hand SNAP will be
able to determine $(\OM^0,w_0)$ to within small errors
$(3\%,5\%)$, and will significantly improve the determination of
the time-variation parameter $w_1$, but only up to $30\%$ at most
(each parameter being marginalized over others) \cite{linder}.
Moreover, the degeneracies among their combinations weaken
substantially the tightness of these bounds.


\section{Concluding remarks}

We have considered a Friedmann-Lema\^{\i}tre-Robertson-Walker
model with time-evolving cosmological term: $\CC=\CC(t)$. The
evolution of the CC is due to quantum effects that can be
described within the Renormalization Group (RG) approach. The RG
scale $\mu$ is identified with the Hubble parameter $H$ at the
corresponding epoch, as first proposed in \cite{JHEPCC1}.
Although the $\bCC$-function for $\CC$ is just proportional to
the fourth power of the masses in the UV regime, the phenomenon
of decoupling in QFT leads to an inverse power suppression by the
heavy masses at low energies (IR regime). Thus, in the present
day universe, one may expect a modified RG equation characterized
by a ``soft decoupling'' behaviour $\bCC\sim H^2
M^2$\,\cite{JHEPCC1,Letter,IRGA03,CCfit}. The effective mass
scale $M$ (\ref{Mdef}) summarizes the presence of the heavy
degrees of freedom. This peculiar form of decoupling can be
envisaged from: \,i) the Appelquist-Carazzone (AC)-decoupling
theorem, \,ii) general covariance of the effective action, and
also from\, iii) the \textit{non}-fine-tuning hypothesis on the
$n=1$ terms of $\bCC$ -- Cf. Eq. (\ref{newRG1})-(\ref{newRG2})--
which insures that the coefficient of the quadratic contribution
$H^2 M^2$ does not vanish. This particular form of decoupling is
a specific feature of the CC because it is of dimension four.
There is no other parameter either in the SM or in GUT models
with such property.

In constructing this semiclassical ``RG-cosmology'' we have
explored the possibility that the heaviest d.o.f. may be
associated to particles having the masses just below the Planck
scale, hence $M$ is of order $M_P$. This assumption is essential
to implement the soft decoupling hypothesis within the $\mu=H$
setting, for the present value of $\,H_0^2 M^2\,$ is just of the
order of the CC. This fact insures a smooth running of the
cosmological term around the present time. In this model the
$\bCC$-function has only one arbitrary parameter $\,\nu\,$
(\ref{nu}) proportional to the dimensionless ratio $M^2/M_P^2$,
and as a result the model has an essential predictive power. In
general we expect $\,|\nu|\ll 1\,$ from phenomenological
considerations, mainly based on the most conservative hypotheses
on nucleosynthesis.

As the variation of the CC is attributed, in this model, to the
``relic'' quantum effects associated to the decoupling of the
heaviest degrees of freedom below the Planck scale, a time
dependence of the CC may be achieved without resorting to scalar
fields mimicking the cosmological term (``quintessence'') or to
modifications of the structure of the SM of the strong and
electroweak interactions and/or of the gravitational
interactions. This proposal, therefore, offers an excellent
opportunity to explore the existence of sub-Planck physics in
direct cosmological experiments, such as SNAP (and the very
high--z SNe Ia data to be obtained with HST). For $\nu\lesssim
0.1$ corrections to some FLRW cosmological parameters become as
large as $50\%$ or more, which could not be missed by these
experiments.

Whether this RG-cosmology can be easily distinguished from
quintessence models requires further considerations along the
lines of previous studies on evolving dark energy
\,\cite{Freeman,Padmanabhan}. The present model, however, can
elude some of the difficulties (related to the degeneracies of the
kinematical and geometrical measurements\,\cite{Padmanabhan}), in
that we predict not only an evolving vacuum energy, but also a
correlated ($\nu$-dependent) departure of the matter-radiation
density from the standard model prediction. From a more
fundamental point of view, the sole fact that our FLRW scenario
with running cosmological constant can compete on the same
footing with quintessence models shows that standard quantum
field theory in curved space-time may contain the necessary
ingredients from which one can build up a time-evolving
cosmological term without need of artificial (``just so'') scalar
fields. Both types of models can be thoroughly checked by the
SNAP and upgraded HST experiments \cite{SNAP}. If these
experiments will detect the redshift dependence of the CC similar
to that which is predicted in our work,  we may suspect that some
relevant physics is going on just below the Planck scale. If, on
the contrary, they unravel a static CC, this may imply the
existence of a desert in the particle spectrum below the Planck
scale, which would be no less noticeable. In this respect let us
not forget that the popular notion of a GUT (perhaps in the form
of string physics) near the Planck scale remains, at the moment,
as a pure (though very much interesting!) theoretical
speculation, which unfortunately is not supported by a single
piece of experimental evidence up to now. Our framework may allow
to explore hints of these theories directly from
astrophysical/cosmological experiments which are just round the
corner. If the results are positive, it would suggest a direct
link between the largest scales in cosmology and the shortest
distances in high energy physics.

\vskip 5mm

\noindent {\large\it Acknowledgments.} Authors are thankful to
E.V. Gorbar, B. Guberina, M. Reuter, H. Stefancic and A.
Starobinsky for fruitful discussions. We thank C. Espa\~na-Bonet
and P. Ruiz-Lapuente for their collaboration in the numerical
analysis. The work of I.Sh. has been supported by the research
grant from FAPEMIG (MG, Brazil) and by the fellowship from CNPq
(Brazil). The work of J.S. has been supported in part by MECYT
and FEDER under project FPA2001-3598. J.S. is grateful to the
organizers of the workshop for the invitation to the conference
and for financial support.



\end{document}